\documentclass[reprint,superscriptaddress,amsmath,amssymb,prl]{revtex4-2}
\usepackage{color}
\usepackage{amsthm,amsmath,amssymb}
\usepackage{mathrsfs}
\usepackage{dsfont}
\usepackage{braket}
\usepackage{graphicx}
\usepackage{dcolumn}
\usepackage{bm}

\begin{document}
\title{Symmetry dictated universal helicity redistribution of Dirac fermions in transport}

\author{Jun-Yin Huang}
\affiliation{Lanzhou Center for Theoretical Physics, Key Laboratory of Theoretical Physics of Gansu Province, and Key Laboratory of Quantum Theory and Applications of MoE, Lanzhou University, Lanzhou, Gansu 730000, China}

\author{Rui-Hua Ni}
\affiliation{Lanzhou Center for Theoretical Physics, Key Laboratory of Theoretical Physics of Gansu Province, and Key Laboratory of Quantum Theory and Applications of MoE, Lanzhou University, Lanzhou, Gansu 730000, China}

\author{Hong-Ya Xu}
\email{Corresponding author: xuhongya@lzu.edu.cn}
\affiliation{Lanzhou Center for Theoretical Physics, Key Laboratory of Theoretical Physics of Gansu Province, and Key Laboratory of Quantum Theory and Applications of MoE, Lanzhou University, Lanzhou, Gansu 730000, China}

\author{Liang Huang}
\email{Corresponding author: huangl@lzu.edu.cn}
\affiliation{Lanzhou Center for Theoretical Physics, Key Laboratory of Theoretical Physics of Gansu Province, and Key Laboratory of Quantum Theory and Applications of MoE, Lanzhou University, Lanzhou, Gansu 730000, China}

\date{\today}

\begin{abstract}
Helicity is a fundamental property of Dirac fermions. Yet, the general rule of how it changes in transport is still lacking. We uncover, theoretically, the universal spinor state transformation and consequently helicity redistribution rule in two cases of transport through potentials of electrostatic and mass types, respectively. The former is dictated by Lorentz boost and its complex counterpart in Klein tunneling regime, which establishes miraculously a unified yet latent connection between helicity, Klein tunneling, and Lorentz boost. The latter is governed by an abstract rotation group we construct, which reduces to SO(2) when acting on the plane of effective mass and momentum. They generate invariant submanifolds, i.e., leaves, that foliate the Hilbert space of Dirac spinors. Our results provide a basis for unified understanding of helicity transport, and may open a new window for exotic helicity-based physics and applications in mesoscopic systems. 
\end{abstract}
\maketitle

{\it Introduction.}---Helicity, the projection of the spin onto the direction of momentum, is known as an intrinsic and measurable property of Dirac fermions in relativistic quantum mechanics, and plays a crucial role in understanding the nature of fundamental particles. For instance, the helicity nature of massless Dirac particles turns out to be responsible for a highly anisotropic tunneling, i.e., chiral tunneling, through an electrostatic potential barrier~\cite{katsnelson2006chiral, zeb2008chiral, tudorovskiy2012chiral,he2013chiral, habib2015chiral}. Flip and conservation of helicity has been recognized as a key issue in addressing scattering of massive Dirac particles by a magnetic monopole~\cite{zwanziger1968exactly, kazama1977scattering} and an Aharonov-Bohm potential~\cite{vera1990helicity,coutinho1994helicity, araujo2001most}. As an independent and exploitable degree of freedom, it attracts attention for electronic applications in mesoscopic systems recently \cite{gu2011cloaking, zhao2015electronic}. However, before stepping forward further, it is crucial to understand the redistribution rules of helicity in typical scattering, tunneling, and transport processes. 

Generally, the transport process can be abstracted as an operation on the Hilbert space of Dirac spinors mapping the initial to the final spinor states characterized by energy and momentum. Yet, the helicity degree of freedom is undetermined as the coefficients of the two helicity components can still take different values, which typically depends on the transport details. One fundamental question is then under commonly encountered transport situations, how does the helicity change? 

By utilizing the (3+1)-D Dirac equation formalism in the single-particle framework, we consider two typical transport 
processes, i.e., through piecewise-constant electrostatic \cite{de2009planar,setare2010klein,de2012relative, bittencourt20152, navarro2020two} or mass potentials \cite{Jackiw_2012, Hunt:2013, shen2017topological, ZhaoRZM:2022, cheng2022monopole, Wang:2022, Alu:2023}. 
For each process, as the potential of the concerned region changes, the final spinor evolves out a curve started from the given initial state. We find, strikingly, that the final spinor depends only on the potential height of this region, i.e., process-independent. Thus the corresponding abstract operation forms a one-parameter transformation group, whose exertion on any initial state generates a one-dimensional invariant submanifold, which forms leaves foliating the Hilbert space of Dirac spinors. 

\begin{figure*}[t]
\includegraphics[width=0.9\linewidth]{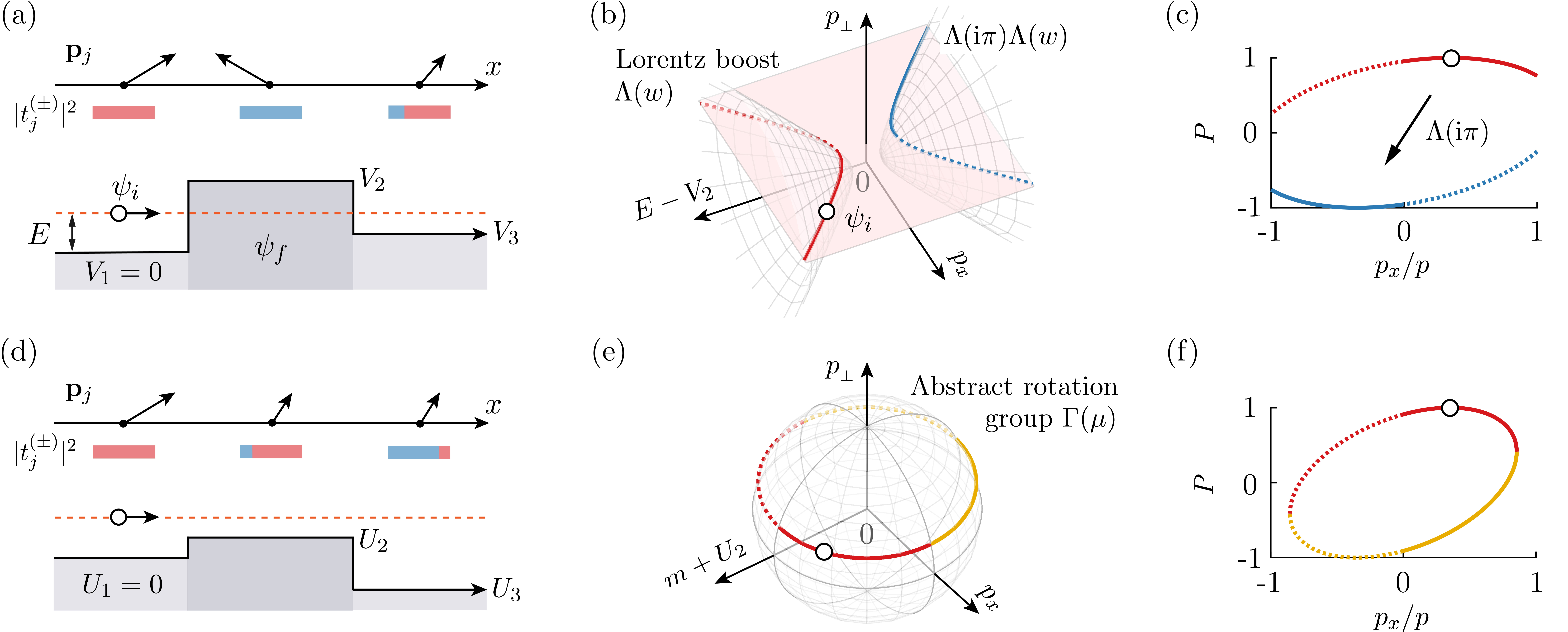}
\caption{\label{fig:1} Schematics of the two transport processes and the rule of spinor state transformations. (a) Transport through a piecewise-constant electrostatic potential, with momentum $\mathbf{p}_j$ and relative helicity components $|t_j^{(\pm)}|^2$ (red for positive, blue for negative) in each region $j$ demonstrated. $\ket{\psi_i}$ is the initial (incident) spinor state with completely positive helicity, and $\ket{\psi_f}$ is the final state in region $2$. (b,c) Visualizing the Lorentz boost $\Lambda(w)$ dictated transformation rule we uncovered from $\ket{\psi_i}$ (hollow circle) to $\ket{\psi_f}$s. Projection of the leaf consisted of all $\ket{\psi_f}$s onto subspaces parameterized by effective energy $E-V_2$, momentum $p_x$ and  $p_{_\bot}$ (b), and by $p_x/p$ and helicity polarization $P$ (c). Solid (dashed) curves are for transmitted (reflected) parts. Red and blue indicate positive and negative effective energy branches, respectively, which are connected by $\Lambda(\mathrm i \pi)$. (d-f) Mass potential case. The transformation rule is now governed by an abstract rotation group $\Gamma(\mu)$ we identified. (e) Projection of the leaf in the ($m+U_2$, $p_x$, $p_{_\bot}$) subspace, and (f) on the ($p_x/p$, $P$) plane. Red and yellow curves are for positive and negative effective mass $m+U_2$, respectively. 
}
\end{figure*}

Our main analytical results are shown in Fig. \ref{fig:1}.
The left panels plot the schematics of the two processes, where the potentials are along the $x$-axis, and the perpendicular momentum component $p_{_\bot}$ is conserved. The final state $\ket{\psi_f}$ is chosen in the middle region to simply include the reflected wavefunction. The middle and right panels show the foliation structures. For transport through an electrostatic potential with a given initial energy $E$ [Fig. \ref{fig:1}(a)], the Hilbert space of spinors can be fully parameterized by the effective energy $E-V_2$, momentum $p_x$ and $p_{_\bot}$, and helicity polarization $P$ \cite{relativephase}. As the potential $V_2$ changes, from any initial state $\ket{\psi_i}$, a one-dimensional leaf grows out in the Hilbert space, i.e., a curve in the four dimensional parameter space. 
Particularly, the projection of this leaf into the ($E-V_2$, $p_x$, $p_{_\bot}$) subspace is a hyperboloid characterized by $E$ [Fig. \ref{fig:1}(b)]. Different $p_{_\bot}$, as determined by the incident angle, corresponds to different sections, i.e., hyperbolic curves. 
The projection onto the ($p_x/p$, $P$) plane is shown in Fig. \ref{fig:1}(c), where $p=\sqrt{p_x^2 + p_{_\bot}^2}$. Surprisingly, the one-parameter transformation group for the spinors and consequently the redistribution of helicity is just the spinor representation of Lorentz boost $\Lambda(w)$ along the $x$-axis. When $V_2$ is large enough, the branch for negative energy states ($E-V_2 < 0$) associated with Klein tunneling emerges, which can be mapped from the original branch by a complex Lorentz boost $\Lambda(\mathrm i \pi)$. This builds miraculously a latent, unified connection between different essential properties, i.e., helicity, Klein tunneling, and Lorentz boost, of Dirac fermions. 
The universal redistribution rule of the helicity is corroborated by extensive numerics in more general cases.

For the mass potential case with initial energy $E$ [Fig. \ref{fig:1}(d)], the effective mass $m+U_2$ plays a similar role of the effective energy. The leaf can then be projected onto the ($p_x$, $m+U_2$) plane, which, in the natural units ($c=1$), is a circle. All these circles for possible $p_{_\bot}$s construct a sphere in the ($m+U_2$, $p_x$, $p_{_\bot}$) space [Fig. \ref{fig:1}(e)]. The projection on the ($p_x$, $P$) plane is shown in Fig. \ref{fig:1}(f). 
As $U_2$ varies, the one-parameter transformation group $\Gamma(\mu)$ acting on the ($p_x$, $m+U_2$) plane can be conveniently described by an SO(2) rotation. Its spinor representation for the abstract spinor state transformation has also been constructed. 

These findings uncover universal redistribution rules of helicity, dictated by symmetries described by Lorentz boost and an abstract rotation group, respectively, in transport processes through electrostatic or mass potentials. 
Our results provide conveniently a modulation strategy of the helicity by simply configuring the electrostatic or mass potential profiles, holding great promise to exploit the new degree of freedom for novel applications. 

{\it Model.}---The transport problem of the (3+1)-D massive Dirac fermions can be described by the Dirac Hamiltonian in the position representation~\cite{greiner2000relativistic}
\begin{equation}
\hat{H}=
\bm{\alpha}\cdot \hat{\mathbf{p}}+\beta [m+U(x)] + V(x),
\end{equation}
where $\bm{\alpha}=\begin{pmatrix}
0&\bm{\sigma} \\
\bm{\sigma}&0
\end{pmatrix}$ with Pauli matrices $\bm{\sigma}=(\sigma_{x},\sigma_{y},\sigma_{z})$, $\hat{\mathbf{p}}$ is the 3-momentum operator,
$\beta=\begin{pmatrix}
\mathds{1}_2&0 \\
0&-\mathds{1}_2
\end{pmatrix}$, $m$ is the Dirac mass, $U$ is the mass potential, and $V$ is the electrostatic potential. Natural units $c=\hbar=e=1$ are employed. For the sake of simplicity, we consider two typical cases, i) $U(x)=0$, and ii) $V(x)=0$.
We exploit a piecewise-constant potential with $n$ regions of width $\{d_j\}$ along the $x$-axis, where $d_1 = d_n = \infty$, and $V_1 = U_1 = 0$, which, in the large $n$ limit, can approach arbitrary continuous potential profile. 

Note that the whole Hamiltonian $\hat{H}$ does not commute with $\hat{h}$. Nevertheless, in each region $j$, the corresponding Hamiltonian $\hat{H}_j$ commutes with the helicity operator \cite{greiner2000relativistic}
\begin{equation}
\hat{h}=\begin{pmatrix}
\bm{\sigma} & 0\\
0 & \bm{\sigma}
\end{pmatrix}\cdot \hat{\mathbf{p}}/p_j,
\end{equation}
where $p_j = \sqrt{(E-V_j)^2-(m+U_j)^2}$, and the incident energy $E>m$. The transmitted wavefunctions (chosen as the common eigenstates of $\hat{H}$, $\hat{\mathbf{p}}$, and $\hat{h}$) are plane wave solutions \cite{strange1998relativistic}
\begin{align}
\psi_{j}^{(h)}=
\begin{pmatrix}
\chi^{(h)}_{j} \\
h k_j\chi^{(h)}_{j}
\end{pmatrix}
 e^{\mathrm i(\mathbf{p}_{j}\cdot \mathbf{x}-E t)},
\label{eq:psi}
\end{align}
where $k_j=p_j/(E-V_j+m+U_j)$, $h=\pm 1$ is the eigenvalue of $\hat{h}$, and
$\chi^{(+)}_j=
[\cos\frac{\theta_j}{2}e^{-\mathrm i\varphi_j/2},
\sin\frac{\theta_j}{2}e^{\mathrm i \varphi_j/2}]^T$, 
$\chi^{(-)}_j=
[-\sin\frac{\theta_j}{2}e^{-\mathrm i\varphi_j/2},
\cos\frac{\theta_j}{2}e^{\mathrm i \varphi_j/2}]^T$ \cite{strange1998relativistic}, 
with $\mathbf{p}_j=(p_{j,x},p_y,p_z) = p_j(\sin\theta_j\cos\varphi_j,\sin\theta_j\sin\varphi_j,\cos\theta_j)$ as $p_y$ and $p_z$ are conserved in different regions \cite{de2009planar, refphi}, 
$\theta_j =\cos^{-1}(p_1\cos\theta_1/p_j)$, 
$\varphi_j =\cos^{-1}(\lambda_j \sqrt{p_j^2-p_y^2-p_z^2}\big/p_j\sin\theta_j)$, 
and $\lambda_j={\rm sgn}(E-V_j)$ is for positive or negative energy states. Due to the rotational symmetry with respect to the $x$-axis, the angle $(\theta_j, \varphi_j)$ can be characterized by a single angle $\varrho_{j}=\cos^{-1}(\sin\theta_j\cos\varphi_{j})$ between $\mathbf{p}_j$ and the positive $x$-axis. For the reflected plane wave $\overline{\psi}_{j}^{(h)}$, one only needs to change $p_{j,x}$ to $-p_{j,x}$ and $\varphi_j$ to $\pi-\varphi_j$. Without loss of generality, we assume that the incident Dirac fermions are completely positively helicity-polarized plane waves.

The wavefunction in each region can be decomposed into the eigenmodes in terms of Eq. (\ref{eq:psi}) and its reflected counterpart, i.e., $\psi_{j} = t_{j}^{(h)}\psi_{j}^{(h)} + r_{j}^{(h)}\overline{\psi}_{j}^{(h)}$.
For transmitted flow, the helicity polarization is defined as \cite{bandyopadhyay2008introduction,eschrig2015spin}
\begin{align}
P_{j}=(J_{j}^{(+)}-J_{j}^{(-)})\big/(J_{j}^{(+)}+J_{j}^{(-)}),
\label{eq:Pj}
\end{align}
where $J_{j}^{(h)}=|t_{j}^{(h)}|^2|\psi_{j}^{(h)\dag} \bm{\alpha}\psi_{j}^{(h)}|$ is the magnitude of the probability current. For reflected flow, $\overline{J}_{j}^{(h)}$ and $\overline{P}_{j}$ can be defined similarly. The transmission and reflection coefficients ($t_{j}^{(h)},r_{j}^{(h)}$) are then numerically calculated via the transfer matrix method by matching them at the boundary of each region \cite{davies1998physics,markos2008wave, cotuaescu2007applying}. The overall transmission probability is then $$T^{(h)} = |t_{n}^{(h)}|^2 k_n\sin\theta_n\cos\varphi_n\big/k_1\sin\theta_1 \cos\varphi_1.$$ 
As a test, by approximating a smooth linear potential with large $n$, our numerical calculation matches perfectly with the analytical result \cite{sauter1931verhalten} [Supplemental Material (SM) Sec. I].

{\it Theory.}---We derive the redistribution rules of helicity in transport as follows. 

{\it Case} i), only consider the electrostatic potential, i.e., for any $j\in [1,n]$, $U_j = 0$.
We have
\begin{align}
\hat{H}_{j}\psi_{j}^{(h)}=\lambda_j |E-V_j|\psi_{j}^{(h)}, \ \hat{h}\psi_{j}^{(h)}=h \psi_{j}^{(h)}.
\end{align}
For $V_j \in (V_-,V_+)$ with $V_{\pm}=E\pm \sqrt{p_y^2+p_z^2+m^2}$, the transmitted part has an imaginary $p_{j,x}$, which is evanescent and no longer has a well-defined helicity. Therefore we only consider $V_j < V_-$ scattering or $V_j > V_+$ Klein tunneling.

For $n=3$, the normalized transmission and reflection coefficients $t_{j}^{(h)}$ ($j=2,3$) and $r_{j}^{(h)}$ ($j=1,2$) can be obtained analytically (SM Sec. II), e.g., 
\begin{align}
\label{eq:M}
\begin{pmatrix}
t_j^{(+)} \\ t_j^{(-)} \\
\end{pmatrix}=|\mathcal{M}|^{-1/2} \begin{pmatrix}
\mathcal{M}_{(+,+)} & \mathcal{M}_{(+,-)} \\ \mathcal{M}_{(-,+)} & \mathcal{M}_{(-,-)} \\
\end{pmatrix}\cdot \begin{pmatrix}
t_1^{(+)} \\ t_1^{(-)} \\
\end{pmatrix},
\end{align}
where $\mathcal{M}_{(h',h)} \equiv \bra{\psi_j^{(h')}}\alpha_x \ket{\psi_1^{(h)}}$ with normalized ket $\ket{\psi_j^{(h)}}=\psi_j^{(h)} /| \psi_j^{(h)}|$. For positively helicity-polarized initial state, $t_1^{(+)}=1$, $t_1^{(-)}=0$. 
Plugging into Eq. (\ref{eq:Pj}), the exact expression of the helicity polarization for the transmitted flow is 
\begin{align}
P_{j}=(\eta_{j}-1)\big/(\eta_{j}+1),
\label{eq:Pjexact}
\end{align}
where
\begin{align}
\eta_{j}=\left|t_j^{(+)}\big/t_j^{(-)}\right|^2 =\left(\frac{k_1+k_j}{k_1-k_j}\right)^2\cdot \frac{1+\cos(\varrho_{1}+\varrho_{j})}{1-\cos(\varrho_{1}+\varrho_{j})}.
\label{eq:etaj}
\end{align}
For reflected flow, one only needs to change $\varrho_j$ to $\pi-\varrho_j$ to get $\overline{P}_j$.

Remarkably, the spinor state $\psi_j^{(h)}$ in region $j$ is determined completely by $V_j$ and $(E,\mathbf{p}_1)$ [Eq. (\ref{eq:psi})], and is irrelevant to the potential in other regions. So do the matrix $\mathcal{M}$ and the helicity polarization $P_{j}$. Regarding the transformation from $\ket{\psi_1}$ to $\ket{\psi_j}$ as an abstract operation, 
it is additive and forms a one-parameter ($V_j$) transformation group (SM Sec. \uppercase\expandafter{\romannumeral3}-A), generating leaves in the spinor Hilbert space. 

\begin{figure} [t]
	\centering
\includegraphics[width=0.7\linewidth]{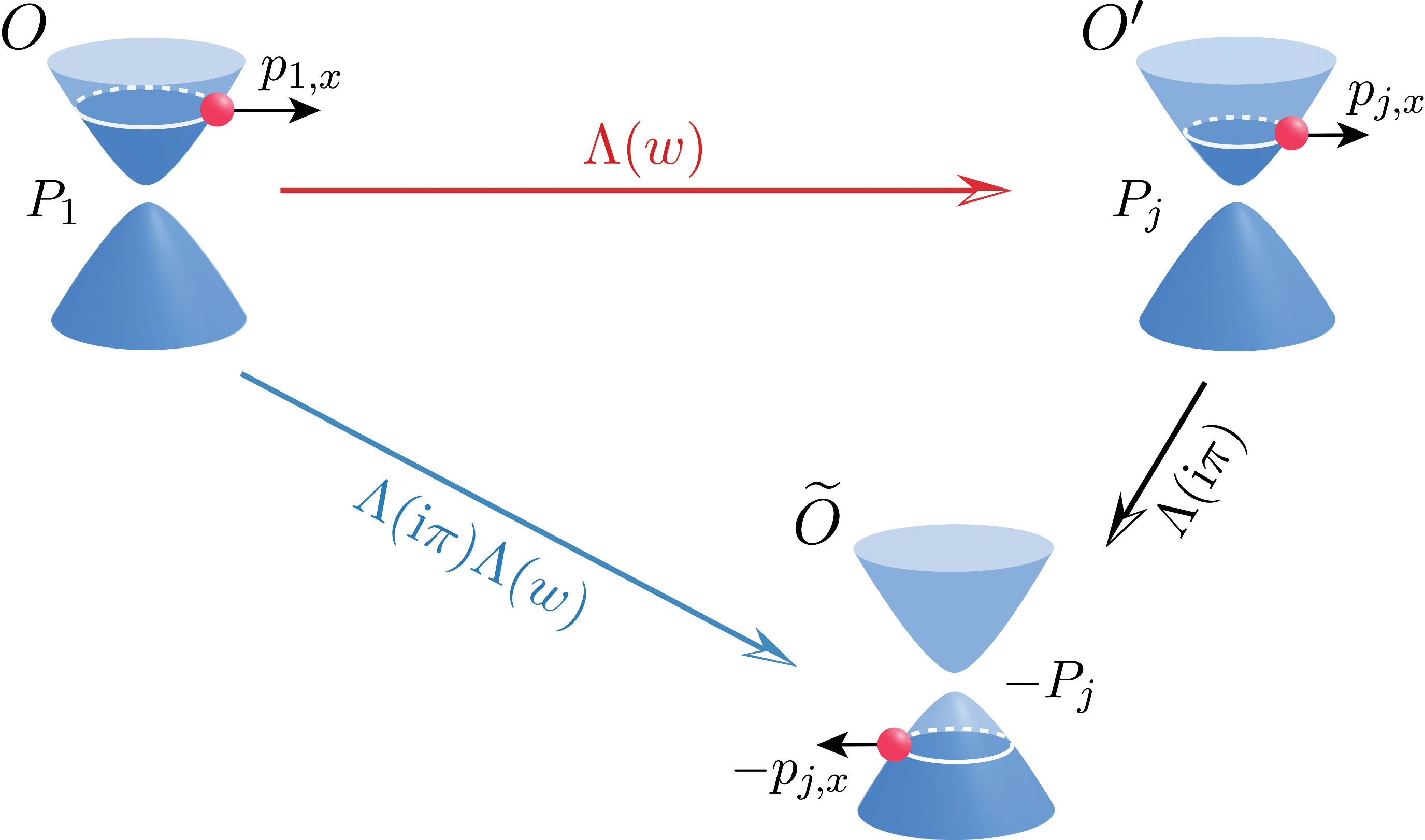}
	\caption{\label{fig:LB} The change of effective energy, momentum, and helicity from $\ket{\psi_1}$ to $\ket{\psi_j}$ in the actual transport problem can be described by a Lorentz boost $\Lambda(w)$ with real rapidity $w$ of the imagined inertial frame transformation from $O$ to $O'$. The Klein tunneling can be described by the combined operation $\Lambda(\mathrm i \pi)\Lambda(w)$, i.e., $O \xrightarrow{\Lambda(w)} O' \xrightarrow{\Lambda(\mathrm i \pi)} \widetilde{O}$.}
\end{figure}

Alternatively, disregard the actual transport process, only consider the two sets of energy and momentum of the initial and final states. Now imagine that they are for the same state but observed in two different inertial reference frames, i.e., $O$ and $O'$ in Fig. \ref{fig:LB}, and are thus connected by a Lorentz boost $\Lambda(w)$ with rapidity $w$ (defined by $\cosh w=(1-v^2)^{-1/2}$ with $v$ being the relative velocity of the two frames), 
\begin{align}
\begin{pmatrix}
  E-V_j \\ p_{j,x}
\end{pmatrix}=\begin{pmatrix}
\cosh w & -\sinh w \\
-\sinh w & \cosh w\\ 
\end{pmatrix}\cdot \begin{pmatrix}
 E \\ p_{1,x}
\end{pmatrix}.
\end{align}
In this regard, the spinor state will be transformed to $\hat{S}\ket{\psi_1}$ with 
\begin{align}
\label{eq:S}
\hat{S}[\Lambda(w)]=\cosh(w/2)\mathds{1}_4-\sinh(w/2)\alpha_x ,
\end{align} 
being the spinor representation of the Lorentz boost \cite{greiner2000relativistic}.
Surprisingly, we find $\hat{S}\ket{\psi_1}$ from the imagined inertial frame transformation is the same as the final state $\ket{\psi_j}$ in the actual transport problem. 
As such, the one-parameter transformation group characterizing the transport process is just the Lorentz boost. Retrospectively, as the electrostatic potential only shifts the energy of the Dirac fermion, the equation structure is preserved, and the corresponding abstract operation is equivalent to the inertial frame transformation (Lorentz boost). This simple, universal behavior unveils yet another concealed attribute of the miraculous Dirac equation. 
 
\begin{figure} [t]
	\centering
         \includegraphics[width=1\linewidth]{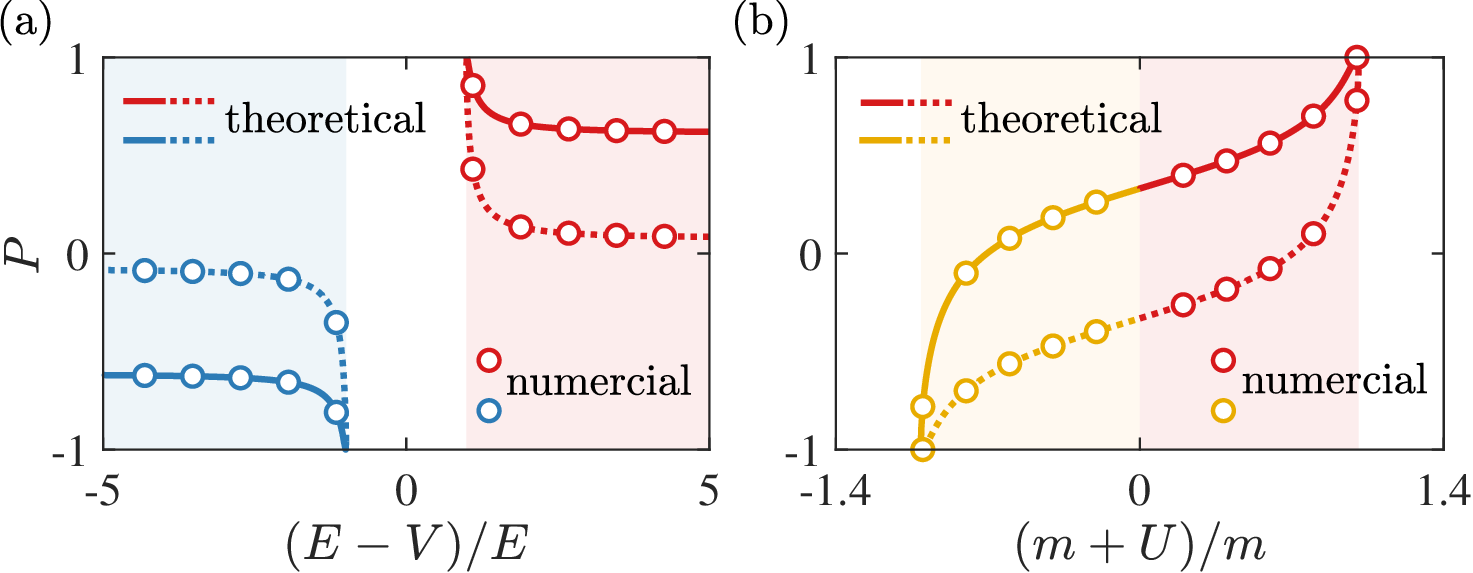}
\caption{\label{fig:num} Plots of helicity polarization for transmitted ($P_{j}$, solid curves) and reflected ($\overline{P}_j$, dashed) waves. (a) vs effective energy $E-V_j$ in units of $E$ for electrostatic potential, and (b) vs effective mass $m+U_j$ in units of $m$ for mass potential. The incident energy and angle: $E=1.08m$, $\varrho_1=2\pi/5$. The curves are from Eqs. (\ref{eq:Pjexact}-\ref{eq:etaj}). The circles are numerical results from the random piecewise-constant potential model with $n=10$ and $j$ a randomly chosen region, e.g., $j=5$.}
\end{figure}

For Klein tunneling ($V_j>V_+$), we will take the convention $p_{j,x}<0$, so that the $x$-component of the group velocity ${v}_{x}=\lambda_j p_{j,x}/|E-V_j|$ is always positive for transmitted part \cite{klein1929streuung,castro2009electronic, nguyen2009tunneling}. 
Conventional Lorentz boost fails to link the positive and negative energy states. By replacing $w$ with $\mathrm i\pi$, the operation $\Lambda(\mathrm i\pi)$ flips the sign of effective energy and $p_x$, as from $O'$ to $\widetilde{O}$ in Fig. \ref{fig:LB}. $\hat{S}[\Lambda(\mathrm i\pi)] = - \mathrm i \alpha_x$ flips the helicity polarization, exactly the same as that in the actual tunneling process. The combined operation from $O$ to $\widetilde{O}$ is then
$\Lambda(\mathrm i \pi)\Lambda(w)$ (SM Sec. \uppercase\expandafter{\romannumeral3}-B), which is traceable \cite{jost1957bemerkung, bergknoff1979structure,thacker1981exact, froggatt1991cpt,lehnert2016cpt}. 
The complex Lorentz boost $\Lambda(\mathrm i\pi)$ enforces an antisymmetry between the helicity polarization and the effective energy $E-V$ [Fig. \ref{fig:num}(a)].

Indeed, for Dirac spinors, introducing $\hat{S}[\Lambda(\mathrm i\pi)]$ is required by the $\mathcal{PCT}$ symmetry. 
The $\mathcal{PCT}$ symmetry interchanges the positive and negative energy states, with $\mathcal{P}=\mathcal{P}_x\mathcal{P}_y\mathcal{P}_z$, $\mathcal{C}$, $\mathcal{T}$ being the parity, charge conjugate, and time-reversal operations. Due to the rotational symmetry with respect to the $x$-axis, $\mathcal{P}_y\mathcal{P}_z =\mathds{1}$, the $\mathcal{PCT}$ symmetry reduces to $\mathcal{P}_x\mathcal{CT}$, 
which equals to $\hat{S}[\Lambda(\mathrm i\pi)]$ disregarding an unobservable phase (SM Sec. \uppercase\expandafter{\romannumeral3}-B). 

The above results do not explicitly depend on the fact of $n=3$, and are valid for arbitrary $n$ regions.

{\it Case} ii), the mass potential. Requiring $p_{j,x}$ to be real leads to $U_j \in (U_-,U_+)$ with $U_{\pm}=-m \pm \sqrt{E^2-p_{y}^2-p_z^2}$. Note that the negative mass potential has been widely used in models of nuclear physics and condensed matter physics \cite{ZhaoRZM:2022, Jackiw_2012, Hunt:2013, shen2017topological, cheng2022monopole, Wang:2022}. Similarly, Eqs. (\ref{eq:M}-\ref{eq:etaj}) have the same form except that now $\psi_j^{(h)}$, $k_j$, and $\varrho_j$ are determined by $U_j$. The corresponding map from $\ket{\psi_1}$ to $\ket{\psi_j}$ also forms a one-parameter transformation group, denoted as $\Gamma(\mu)$ (SM Sec. IV). In the plane of effective mass $m+U_j$ and $x$-momentum, the action of $\Gamma(\mu)$ reduces to SO(2) rotation 
\begin{align}
\begin{pmatrix}
  m+U_j \\ p_{j,x}
\end{pmatrix}=\begin{pmatrix}
\cos\mu & -\sin\mu \\
\sin\mu & \cos\mu \\ 
\end{pmatrix}\cdot \begin{pmatrix}
 m \\ p_{1,x}
\end{pmatrix},
\end{align}
where $\mu$ is an angle parameter determined by $U_j$. The transformation of the spinor states from $\ket{\psi_1}$ to $\ket{\psi_j}$ in the actual transport process can be constructed as $\ket{\psi_j} = \hat{S}_m\ket{\psi_1}$, with
\begin{align}
\hat{S}_m[\Gamma(\mu)] = \cos(\mu/2)\mathds{1}_4-\sin(\mu/2)\beta\alpha_x
\end{align}
being the spinor representation of $\Gamma(\mu)$. Thus the redistribution rule of the helicity is derived, and is completely different from that due to electrostatic potentials. To the best of our knowledge, this group $\Gamma(\mu)$ and its spinor representation are unknown before. 

{\it Numerical verification.}---To be as general as possible, we consider a piecewise-constant potential model with $n=10$ and random widths and potential heights for each region: $d_j \in (0,1/m)$, $V_j$ or $U_j\in(-10m,10m)$. 
We then systematically vary the potential value in a randomly chosen region, say, $j=5$, and plot the helicity polarization of the transmitted and reflected waves in Fig. \ref{fig:num} as symbols. They agree with the theory well for both of the electrostatic and mass potentials. 
The numerical simulation corroborates clearly that the helicity redistribution only depends on the potential in this region, but is independent of the transport processes before or after this region. 

\begin{figure} [tbp]
	\centering
\includegraphics[width=\linewidth]{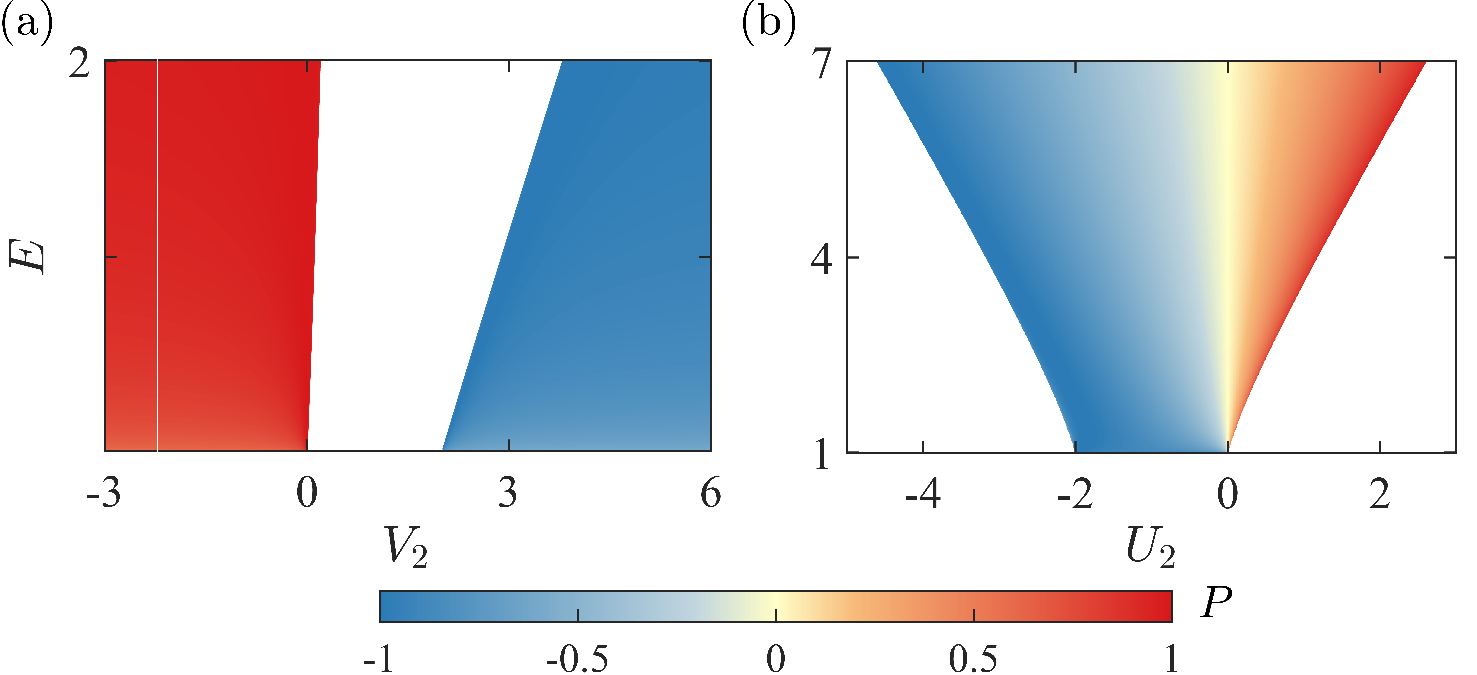}
\caption{\label{fig:apps} Density maps for helicity polarization in the output region ($j=n=2$) through step potentials. (a) vs incident energy $E$ and electrostatic potential $V_2$, and (b) vs $E$ and mass potential $U_2$, all in units of $m$. The incident angle $\varrho_1=\pi/3$. The helicity polarization of the incident waves is $P=1$ for (a) and $P=0$ ($t_1^{(+)}=1$ and $t_1^{(-)}=\mathrm i$) for (b).}
\end{figure}

{\it Potential applications.}---Figure \ref{fig:apps} demonstrates the variation of the helicity polarization $P$ vs the key parameters. For electrostatic potential, $P$ can be reversed from $1$ (incident) to $-1$ (output). For mass potential, with a non-polarized incident wave of $P=0$, the output can be modulated to either $1$ or $-1$. That is, complete polarization of helicity can be generated from completely non-polarized incident waves by applying a properly adjusted mass potential. 
Due to the inherent connection to spin polarization, the helicity, with the unveiled redistribution rule, may lead to appealing applications for spin-based electronics. 

{\it Discussion and conclusion.}---
Another transport setup is through magnetic barriers. In this case, the helicity is conserved \cite{de2007magnetic, wu2010electron, johnson1997hybrid, nogaret2000resistance} as it commutes with the (3+1)-D Dirac Hamiltonian in a static magnetic field \cite{goldhaber1977dirac, vera1990helicity, andreev2018longitudinal, wang2021helical}. Thus the redistribution rule is trivial. 

Helicity is one of the most fundamental and intriguing properties of Dirac fermions. Due to the complexity of its combined nature of spin and momentum, a unified understanding of how it changes in transport is still lacking. 
We tackle this issue and have uncovered the universal helicity redistribution rule, from which process-independent helicity polarizations arise. 
The underlying principle of the universality is unravelled by abstracting the physical transport of helicity into imagined transforming operations on the spinor states. 
In such way, we identify one-parameter transformation groups as well as their spinor representations, which are described by the Lorentz boost $\Lambda(w)$ for electrostatic potentials, and the abstract rotation group $\Gamma(\mu)$ for mass potentials. 
This explicitly reveals an unknown connection between the physical rule of helicity in transport that is highly nontrivial and the underlying symmetries of the Dirac equation. 
Our results not only are fundamental to the understanding of relativistic quantum scattering, tunneling, and transport, but also offer promises of generating desired helicity polarization for exotic applications in mesoscopic systems \cite{zhao2015electronic, wang2016electronic, Gogin_2022, Gogin_2022_2}. 

\begin{acknowledgments}
{\it Acknowledgments.}---
We thank Profs. Yong-Shi Wu, Zhong-Zhou Ren, Li-Sheng Geng for insightful discussions. 
This work was supported by NSFC under Grants No. 12175090, No. 12105125, No. 11775101, and No. 12247101, and by the 111 Project under Grant No. B20063. 
\end{acknowledgments}

\bibliography{paper}

\providecommand{\noopsort}[1]{}\providecommand{\singleletter}[1]{#1}%
\begin{thebibliography}{55}%
\makeatletter
\providecommand \@ifxundefined [1]{%
 \@ifx{#1\undefined}
}%
\providecommand \@ifnum [1]{%
 \ifnum #1\expandafter \@firstoftwo
 \else \expandafter \@secondoftwo
 \fi
}%
\providecommand \@ifx [1]{%
 \ifx #1\expandafter \@firstoftwo
 \else \expandafter \@secondoftwo
 \fi
}%
\providecommand \natexlab [1]{#1}%
\providecommand \enquote  [1]{``#1''}%
\providecommand \bibnamefont  [1]{#1}%
\providecommand \bibfnamefont [1]{#1}%
\providecommand \citenamefont [1]{#1}%
\providecommand \href@noop [0]{\@secondoftwo}%
\providecommand \href [0]{\begingroup \@sanitize@url \@href}%
\providecommand \@href[1]{\@@startlink{#1}\@@href}%
\providecommand \@@href[1]{\endgroup#1\@@endlink}%
\providecommand \@sanitize@url [0]{\catcode `\\12\catcode `\$12\catcode
  `\&12\catcode `\#12\catcode `\^12\catcode `\_12\catcode `\%12\relax}%
\providecommand \@@startlink[1]{}%
\providecommand \@@endlink[0]{}%
\providecommand \url  [0]{\begingroup\@sanitize@url \@url }%
\providecommand \@url [1]{\endgroup\@href {#1}{\urlprefix }}%
\providecommand \urlprefix  [0]{URL }%
\providecommand \Eprint [0]{\href }%
\providecommand \doibase [0]{https://doi.org/}%
\providecommand \selectlanguage [0]{\@gobble}%
\providecommand \bibinfo  [0]{\@secondoftwo}%
\providecommand \bibfield  [0]{\@secondoftwo}%
\providecommand \translation [1]{[#1]}%
\providecommand \BibitemOpen [0]{}%
\providecommand \bibitemStop [0]{}%
\providecommand \bibitemNoStop [0]{.\EOS\space}%
\providecommand \EOS [0]{\spacefactor3000\relax}%
\providecommand \BibitemShut  [1]{\csname bibitem#1\endcsname}%
\let\auto@bib@innerbib\@empty
\bibitem [{\citenamefont {Katsnelson}\ \emph {et~al.}(2006)\citenamefont
  {Katsnelson}, \citenamefont {Novoselov},\ and\ \citenamefont
  {Geim}}]{katsnelson2006chiral}%
  \BibitemOpen
  \bibfield  {author} {\bibinfo {author} {\bibfnamefont {M.}~\bibnamefont
  {Katsnelson}}, \bibinfo {author} {\bibfnamefont {K.}~\bibnamefont
  {Novoselov}},\ and\ \bibinfo {author} {\bibfnamefont {A.}~\bibnamefont
  {Geim}},\ }\bibfield  {title} {\bibinfo {title} {Chiral tunnelling and the
  {K}lein paradox in graphene},\ }\href@noop {} {\bibfield  {journal} {\bibinfo
   {journal} {Nat. Phys.}\ }\textbf {\bibinfo {volume} {2}},\ \bibinfo {pages}
  {620} (\bibinfo {year} {2006})}\BibitemShut {NoStop}%
\bibitem [{\citenamefont {Zeb}\ \emph {et~al.}(2008)\citenamefont {Zeb},
  \citenamefont {Sabeeh},\ and\ \citenamefont {Tahir}}]{zeb2008chiral}%
  \BibitemOpen
  \bibfield  {author} {\bibinfo {author} {\bibfnamefont {M.~A.}\ \bibnamefont
  {Zeb}}, \bibinfo {author} {\bibfnamefont {K.}~\bibnamefont {Sabeeh}},\ and\
  \bibinfo {author} {\bibfnamefont {M.}~\bibnamefont {Tahir}},\ }\bibfield
  {title} {\bibinfo {title} {Chiral tunneling through a time-periodic potential
  in monolayer graphene},\ }\href@noop {} {\bibfield  {journal} {\bibinfo
  {journal} {Phys. Rev. B}\ }\textbf {\bibinfo {volume} {78}},\ \bibinfo
  {pages} {165420} (\bibinfo {year} {2008})}\BibitemShut {NoStop}%
\bibitem [{\citenamefont {Tudorovskiy}\ \emph {et~al.}(2012)\citenamefont
  {Tudorovskiy}, \citenamefont {Reijnders},\ and\ \citenamefont
  {Katsnelson}}]{tudorovskiy2012chiral}%
  \BibitemOpen
  \bibfield  {author} {\bibinfo {author} {\bibfnamefont {T.}~\bibnamefont
  {Tudorovskiy}}, \bibinfo {author} {\bibfnamefont {K.~J.~A.}\ \bibnamefont
  {Reijnders}},\ and\ \bibinfo {author} {\bibfnamefont {M.~I.}\ \bibnamefont
  {Katsnelson}},\ }\bibfield  {title} {\bibinfo {title} {Chiral tunneling in
  single-layer and bilayer graphene},\ }\href@noop {} {\bibfield  {journal}
  {\bibinfo  {journal} {Phys. Scr.}\ }\textbf {\bibinfo {volume} {T146}},\
  \bibinfo {pages} {014010} (\bibinfo {year} {2012})}\BibitemShut {NoStop}%
\bibitem [{\citenamefont {He}\ \emph {et~al.}(2013)\citenamefont {He},
  \citenamefont {Chu},\ and\ \citenamefont {He}}]{he2013chiral}%
  \BibitemOpen
  \bibfield  {author} {\bibinfo {author} {\bibfnamefont {W.-Y.}\ \bibnamefont
  {He}}, \bibinfo {author} {\bibfnamefont {Z.-D.}\ \bibnamefont {Chu}},\ and\
  \bibinfo {author} {\bibfnamefont {L.}~\bibnamefont {He}},\ }\bibfield
  {title} {\bibinfo {title} {Chiral tunneling in a twisted graphene bilayer},\
  }\href@noop {} {\bibfield  {journal} {\bibinfo  {journal} {Phys. Rev. Lett.}\
  }\textbf {\bibinfo {volume} {111}},\ \bibinfo {pages} {066803} (\bibinfo
  {year} {2013})}\BibitemShut {NoStop}%
\bibitem [{\citenamefont {Habib}\ \emph {et~al.}(2015)\citenamefont {Habib},
  \citenamefont {Sajjad},\ and\ \citenamefont {Ghosh}}]{habib2015chiral}%
  \BibitemOpen
  \bibfield  {author} {\bibinfo {author} {\bibfnamefont {K.~M.~M.}\
  \bibnamefont {Habib}}, \bibinfo {author} {\bibfnamefont {R.~N.}\ \bibnamefont
  {Sajjad}},\ and\ \bibinfo {author} {\bibfnamefont {A.~W.}\ \bibnamefont
  {Ghosh}},\ }\bibfield  {title} {\bibinfo {title} {Chiral tunneling of
  topological states: {T}owards the efficient generation of spin current using
  spin-momentum locking},\ }\href@noop {} {\bibfield  {journal} {\bibinfo
  {journal} {Phys. Rev. Lett.}\ }\textbf {\bibinfo {volume} {114}},\ \bibinfo
  {pages} {176801} (\bibinfo {year} {2015})}\BibitemShut {NoStop}%
\bibitem [{\citenamefont {Zwanziger}(1968)}]{zwanziger1968exactly}%
  \BibitemOpen
  \bibfield  {author} {\bibinfo {author} {\bibfnamefont {D.}~\bibnamefont
  {Zwanziger}},\ }\bibfield  {title} {\bibinfo {title} {Exactly soluble
  nonrelativistic model of particles with both electric and magnetic charges},\
  }\href@noop {} {\bibfield  {journal} {\bibinfo  {journal} {Phys. Rev.}\
  }\textbf {\bibinfo {volume} {176}},\ \bibinfo {pages} {1480} (\bibinfo {year}
  {1968})}\BibitemShut {NoStop}%
\bibitem [{\citenamefont {Kazama}\ \emph {et~al.}(1977)\citenamefont {Kazama},
  \citenamefont {Yang},\ and\ \citenamefont
  {Goldhaber}}]{kazama1977scattering}%
  \BibitemOpen
  \bibfield  {author} {\bibinfo {author} {\bibfnamefont {Y.}~\bibnamefont
  {Kazama}}, \bibinfo {author} {\bibfnamefont {C.~N.}\ \bibnamefont {Yang}},\
  and\ \bibinfo {author} {\bibfnamefont {A.~S.}\ \bibnamefont {Goldhaber}},\
  }\bibfield  {title} {\bibinfo {title} {Scattering of a {D}irac particle with
  charge {Z}e by a fixed magnetic monopole},\ }\href@noop {} {\bibfield
  {journal} {\bibinfo  {journal} {Phys. Rev. D}\ }\textbf {\bibinfo {volume}
  {15}},\ \bibinfo {pages} {2287} (\bibinfo {year} {1977})}\BibitemShut
  {NoStop}%
\bibitem [{\citenamefont {Vera}\ and\ \citenamefont
  {Schmidt}(1990)}]{vera1990helicity}%
  \BibitemOpen
  \bibfield  {author} {\bibinfo {author} {\bibfnamefont {F.}~\bibnamefont
  {Vera}}\ and\ \bibinfo {author} {\bibfnamefont {I.}~\bibnamefont {Schmidt}},\
  }\bibfield  {title} {\bibinfo {title} {Helicity and the {A}haronov-{B}ohm
  effect},\ }\href@noop {} {\bibfield  {journal} {\bibinfo  {journal} {Phys.
  Rev. D}\ }\textbf {\bibinfo {volume} {42}},\ \bibinfo {pages} {3591}
  (\bibinfo {year} {1990})}\BibitemShut {NoStop}%
\bibitem [{\citenamefont {Coutinho}\ and\ \citenamefont
  {Perez}(1994)}]{coutinho1994helicity}%
  \BibitemOpen
  \bibfield  {author} {\bibinfo {author} {\bibfnamefont {F.~A.~B.}\
  \bibnamefont {Coutinho}}\ and\ \bibinfo {author} {\bibfnamefont {J.~F.}\
  \bibnamefont {Perez}},\ }\bibfield  {title} {\bibinfo {title} {Helicity
  conservation in the {A}haronov-{B}ohm scattering of {D}irac particles},\
  }\href@noop {} {\bibfield  {journal} {\bibinfo  {journal} {Phys. Rev. D}\
  }\textbf {\bibinfo {volume} {49}},\ \bibinfo {pages} {2092} (\bibinfo {year}
  {1994})}\BibitemShut {NoStop}%
\bibitem [{\citenamefont {Araujo}\ \emph {et~al.}(2001)\citenamefont {Araujo},
  \citenamefont {Coutinho},\ and\ \citenamefont {Perez}}]{araujo2001most}%
  \BibitemOpen
  \bibfield  {author} {\bibinfo {author} {\bibfnamefont {V.~S.}\ \bibnamefont
  {Araujo}}, \bibinfo {author} {\bibfnamefont {F.~A.~B.}\ \bibnamefont
  {Coutinho}},\ and\ \bibinfo {author} {\bibfnamefont {J.~F.}\ \bibnamefont
  {Perez}},\ }\bibfield  {title} {\bibinfo {title} {On the most general
  boundary conditions for the {A}haronov-{B}ohm scattering of a {D}irac
  particle: helicity and {A}haronov-{B}ohm symmetry conservation},\ }\href@noop
  {} {\bibfield  {journal} {\bibinfo  {journal} {J. Phys. A}\ }\textbf
  {\bibinfo {volume} {34}},\ \bibinfo {pages} {8859} (\bibinfo {year}
  {2001})}\BibitemShut {NoStop}%
\bibitem [{\citenamefont {Gu}\ \emph {et~al.}(2011)\citenamefont {Gu},
  \citenamefont {Rudner},\ and\ \citenamefont {Levitov}}]{gu2011cloaking}%
  \BibitemOpen
  \bibfield  {author} {\bibinfo {author} {\bibfnamefont {N.}~\bibnamefont
  {Gu}}, \bibinfo {author} {\bibfnamefont {M.}~\bibnamefont {Rudner}},\ and\
  \bibinfo {author} {\bibfnamefont {L.}~\bibnamefont {Levitov}},\ }\bibfield
  {title} {\bibinfo {title} {Chirality-assisted electronic cloaking of confined
  states in bilayer graphene},\ }\href@noop {} {\bibfield  {journal} {\bibinfo
  {journal} {Phys. Rev. Lett.}\ }\textbf {\bibinfo {volume} {107}},\ \bibinfo
  {pages} {156603} (\bibinfo {year} {2011})}\BibitemShut {NoStop}%
\bibitem [{\citenamefont {Zhao}\ \emph {et~al.}(2015)\citenamefont {Zhao},
  \citenamefont {Wang}, \citenamefont {Liu}, \citenamefont {Xu}, \citenamefont
  {Gu}, \citenamefont {Xue},\ and\ \citenamefont {Duan}}]{zhao2015electronic}%
  \BibitemOpen
  \bibfield  {author} {\bibinfo {author} {\bibfnamefont {L.}~\bibnamefont
  {Zhao}}, \bibinfo {author} {\bibfnamefont {J.}~\bibnamefont {Wang}}, \bibinfo
  {author} {\bibfnamefont {J.}~\bibnamefont {Liu}}, \bibinfo {author}
  {\bibfnamefont {Y.}~\bibnamefont {Xu}}, \bibinfo {author} {\bibfnamefont
  {B.-L.}\ \bibnamefont {Gu}}, \bibinfo {author} {\bibfnamefont {Q.-K.}\
  \bibnamefont {Xue}},\ and\ \bibinfo {author} {\bibfnamefont {W.}~\bibnamefont
  {Duan}},\ }\bibfield  {title} {\bibinfo {title} {Electronic analog of chiral
  metamaterial: {H}elicity-resolved filtering and focusing of {D}irac fermions
  in thin films of topological materials},\ }\href@noop {} {\bibfield
  {journal} {\bibinfo  {journal} {Phys. Rev. B}\ }\textbf {\bibinfo {volume}
  {92}},\ \bibinfo {pages} {041408(R)} (\bibinfo {year} {2015})}\BibitemShut
  {NoStop}%
\bibitem [{\citenamefont {De~Leo}\ and\ \citenamefont
  {Rotelli}(2009)}]{de2009planar}%
  \BibitemOpen
  \bibfield  {author} {\bibinfo {author} {\bibfnamefont {S.}~\bibnamefont
  {De~Leo}}\ and\ \bibinfo {author} {\bibfnamefont {P.}~\bibnamefont
  {Rotelli}},\ }\bibfield  {title} {\bibinfo {title} {Planar {D}irac
  diffusion},\ }\href@noop {} {\bibfield  {journal} {\bibinfo  {journal} {Eur.
  Phys. J. C}\ }\textbf {\bibinfo {volume} {63}},\ \bibinfo {pages} {157}
  (\bibinfo {year} {2009})}\BibitemShut {NoStop}%
\bibitem [{\citenamefont {Setare}\ and\ \citenamefont
  {Jahani}(2010)}]{setare2010klein}%
  \BibitemOpen
  \bibfield  {author} {\bibinfo {author} {\bibfnamefont {M.}~\bibnamefont
  {Setare}}\ and\ \bibinfo {author} {\bibfnamefont {D.}~\bibnamefont
  {Jahani}},\ }\bibfield  {title} {\bibinfo {title} {Klein tunneling of massive
  {D}irac fermions in single-layer graphene},\ }\href@noop {} {\bibfield
  {journal} {\bibinfo  {journal} {Phys. B: Condens. Matter}\ }\textbf {\bibinfo
  {volume} {405}},\ \bibinfo {pages} {1433} (\bibinfo {year}
  {2010})}\BibitemShut {NoStop}%
\bibitem [{\citenamefont {De~Leo}\ and\ \citenamefont
  {Rotelli}(2012)}]{de2012relative}%
  \BibitemOpen
  \bibfield  {author} {\bibinfo {author} {\bibfnamefont {S.}~\bibnamefont
  {De~Leo}}\ and\ \bibinfo {author} {\bibfnamefont {P.}~\bibnamefont
  {Rotelli}},\ }\bibfield  {title} {\bibinfo {title} {Relative helicity phases
  in planar {D}irac scattering},\ }\href@noop {} {\bibfield  {journal}
  {\bibinfo  {journal} {Phys. Rev. A}\ }\textbf {\bibinfo {volume} {86}},\
  \bibinfo {pages} {032113} (\bibinfo {year} {2012})}\BibitemShut {NoStop}%
\bibitem [{\citenamefont {Bittencourt}\ \emph {et~al.}(2015)\citenamefont
  {Bittencourt}, \citenamefont {Mizrahi},\ and\ \citenamefont
  {Bernardini}}]{bittencourt20152}%
  \BibitemOpen
  \bibfield  {author} {\bibinfo {author} {\bibfnamefont {V.~A. S.~V.}\
  \bibnamefont {Bittencourt}}, \bibinfo {author} {\bibfnamefont {S.~S.}\
  \bibnamefont {Mizrahi}},\ and\ \bibinfo {author} {\bibfnamefont {A.~E.}\
  \bibnamefont {Bernardini}},\ }\bibfield  {title} {\bibinfo {title} {S{U} (2)
  $\otimes$ {SU} (2) bi-spinor structure entanglement induced by a step
  potential barrier scattering in two-dimensions},\ }\href@noop {} {\bibfield
  {journal} {\bibinfo  {journal} {Ann. Phys.}\ }\textbf {\bibinfo {volume}
  {355}},\ \bibinfo {pages} {35} (\bibinfo {year} {2015})}\BibitemShut
  {NoStop}%
\bibitem [{\citenamefont {Navarro-Giraldo}\ and\ \citenamefont
  {Quimbay}(2020)}]{navarro2020two}%
  \BibitemOpen
  \bibfield  {author} {\bibinfo {author} {\bibfnamefont {J.}~\bibnamefont
  {Navarro-Giraldo}}\ and\ \bibinfo {author} {\bibfnamefont {C.}~\bibnamefont
  {Quimbay}},\ }\bibfield  {title} {\bibinfo {title} {Two-dimensional {K}lein
  tunneling for massive {D}irac fermions with a defined helicity},\ }\href@noop
  {} {\bibfield  {journal} {\bibinfo  {journal} {Ann. Phys.}\ }\textbf
  {\bibinfo {volume} {412}},\ \bibinfo {pages} {168022} (\bibinfo {year}
  {2020})}\BibitemShut {NoStop}%
\bibitem [{\citenamefont {Jackiw}(2012)}]{Jackiw_2012}%
  \BibitemOpen
  \bibfield  {author} {\bibinfo {author} {\bibfnamefont {R.}~\bibnamefont
  {Jackiw}},\ }\bibfield  {title} {\bibinfo {title} {Fractional and {Majorana}
  fermions: the physics of zero-energy modes},\ }\href
  {https://doi.org/10.1088/0031-8949/2012/T146/014005} {\bibfield  {journal}
  {\bibinfo  {journal} {Phys. Scripta}\ }\textbf {\bibinfo {volume} {2012}},\
  \bibinfo {pages} {014005} (\bibinfo {year} {2012})}\BibitemShut {NoStop}%
\bibitem [{\citenamefont {Hunt}\ \emph {et~al.}(2013)\citenamefont {Hunt},
  \citenamefont {Sanchez-Yamagishi}, \citenamefont {Young}, \citenamefont
  {Yankowitz}, \citenamefont {LeRoy}, \citenamefont {Watanabe}, \citenamefont
  {Taniguchi}, \citenamefont {Moon}, \citenamefont {Koshino}, \citenamefont
  {Jarillo-Herrero},\ and\ \citenamefont {Ashoori}}]{Hunt:2013}%
  \BibitemOpen
  \bibfield  {author} {\bibinfo {author} {\bibfnamefont {B.}~\bibnamefont
  {Hunt}}, \bibinfo {author} {\bibfnamefont {J.~D.}\ \bibnamefont
  {Sanchez-Yamagishi}}, \bibinfo {author} {\bibfnamefont {A.~F.}\ \bibnamefont
  {Young}}, \bibinfo {author} {\bibfnamefont {M.}~\bibnamefont {Yankowitz}},
  \bibinfo {author} {\bibfnamefont {B.~J.}\ \bibnamefont {LeRoy}}, \bibinfo
  {author} {\bibfnamefont {K.}~\bibnamefont {Watanabe}}, \bibinfo {author}
  {\bibfnamefont {T.}~\bibnamefont {Taniguchi}}, \bibinfo {author}
  {\bibfnamefont {P.}~\bibnamefont {Moon}}, \bibinfo {author} {\bibfnamefont
  {M.}~\bibnamefont {Koshino}}, \bibinfo {author} {\bibfnamefont
  {P.}~\bibnamefont {Jarillo-Herrero}},\ and\ \bibinfo {author} {\bibfnamefont
  {R.~C.}\ \bibnamefont {Ashoori}},\ }\bibfield  {title} {\bibinfo {title}
  {Massive {Dirac} fermions and {Hofstadter} butterfly in a van der {Waals}
  heterostructure},\ }\href {https://doi.org/10.1126/science.1237240}
  {\bibfield  {journal} {\bibinfo  {journal} {Science}\ }\textbf {\bibinfo
  {volume} {340}},\ \bibinfo {pages} {1427} (\bibinfo {year}
  {2013})}\BibitemShut {NoStop}%
\bibitem [{\citenamefont {Shen}(2017)}]{shen2017topological}%
  \BibitemOpen
  \bibfield  {author} {\bibinfo {author} {\bibfnamefont {S.-Q.}\ \bibnamefont
  {Shen}},\ }\href@noop {} {\emph {\bibinfo {title} {Topological Insulators:
  Dirac Equation in Condensed Matter}}},\ Springer Series in Solid-State
  Sciences\ (\bibinfo  {publisher} {Springer Singapore},\ \bibinfo {year}
  {2017})\BibitemShut {NoStop}%
\bibitem [{\citenamefont {Zhao}\ \emph {et~al.}(2022)\citenamefont {Zhao},
  \citenamefont {Ren}, \citenamefont {Zhao},\ and\ \citenamefont
  {Meng}}]{ZhaoRZM:2022}%
  \BibitemOpen
  \bibfield  {author} {\bibinfo {author} {\bibfnamefont {Q.}~\bibnamefont
  {Zhao}}, \bibinfo {author} {\bibfnamefont {Z.}~\bibnamefont {Ren}}, \bibinfo
  {author} {\bibfnamefont {P.}~\bibnamefont {Zhao}},\ and\ \bibinfo {author}
  {\bibfnamefont {J.}~\bibnamefont {Meng}},\ }\bibfield  {title} {\bibinfo
  {title} {Covariant density functional theory with localized exchange terms},\
  }\href@noop {} {\bibfield  {journal} {\bibinfo  {journal} {Phys. Rev. C}\
  }\textbf {\bibinfo {volume} {106}},\ \bibinfo {pages} {034315} (\bibinfo
  {year} {2022})}\BibitemShut {NoStop}%
\bibitem [{\citenamefont {Cheng}\ \emph {et~al.}(2022)\citenamefont {Cheng},
  \citenamefont {Yang}, \citenamefont {Wang},\ and\ \citenamefont
  {Lu}}]{cheng2022monopole}%
  \BibitemOpen
  \bibfield  {author} {\bibinfo {author} {\bibfnamefont {H.}~\bibnamefont
  {Cheng}}, \bibinfo {author} {\bibfnamefont {J.}~\bibnamefont {Yang}},
  \bibinfo {author} {\bibfnamefont {Z.}~\bibnamefont {Wang}},\ and\ \bibinfo
  {author} {\bibfnamefont {L.}~\bibnamefont {Lu}},\ }\href@noop {} {\bibinfo
  {title} {Monopole topological resonators}} (\bibinfo {year} {2022}),\ \Eprint
  {https://arxiv.org/abs/2210.09056} {arXiv:2210.09056 [cond-mat.mes-hall]}
  \BibitemShut {NoStop}%
\bibitem [{\citenamefont {Wang}\ \emph {et~al.}(2022)\citenamefont {Wang},
  \citenamefont {Ma}, \citenamefont {Liu}, \citenamefont {Zhang}, \citenamefont
  {Zhang}, \citenamefont {Ke}, \citenamefont {Liu},\ and\ \citenamefont
  {Chan}}]{Wang:2022}%
  \BibitemOpen
  \bibfield  {author} {\bibinfo {author} {\bibfnamefont {M.}~\bibnamefont
  {Wang}}, \bibinfo {author} {\bibfnamefont {Q.}~\bibnamefont {Ma}}, \bibinfo
  {author} {\bibfnamefont {S.}~\bibnamefont {Liu}}, \bibinfo {author}
  {\bibfnamefont {R.-Y.}\ \bibnamefont {Zhang}}, \bibinfo {author}
  {\bibfnamefont {L.}~\bibnamefont {Zhang}}, \bibinfo {author} {\bibfnamefont
  {M.}~\bibnamefont {Ke}}, \bibinfo {author} {\bibfnamefont {Z.}~\bibnamefont
  {Liu}},\ and\ \bibinfo {author} {\bibfnamefont {C.~T.}\ \bibnamefont
  {Chan}},\ }\bibfield  {title} {\bibinfo {title} {Observation of boundary
  induced chiral anomaly bulk states and their transport properties},\ }\href
  {https://doi.org/10.1038/s41467-022-33447-x} {\bibfield  {journal} {\bibinfo
  {journal} {Nat. Commun.}\ }\textbf {\bibinfo {volume} {13}},\ \bibinfo
  {pages} {5916} (\bibinfo {year} {2022})}\BibitemShut {NoStop}%
\bibitem [{\citenamefont {Chen}\ \emph {et~al.}(2023)\citenamefont {Chen},
  \citenamefont {Komissarenko}, \citenamefont {Smirnova}, \citenamefont
  {Vakulenko}, \citenamefont {Kiriushechkina}, \citenamefont {Volkovskaya},
  \citenamefont {Guddala}, \citenamefont {Menon}, \citenamefont {Alù},\ and\
  \citenamefont {Khanikaev}}]{Alu:2023}%
  \BibitemOpen
  \bibfield  {author} {\bibinfo {author} {\bibfnamefont {K.}~\bibnamefont
  {Chen}}, \bibinfo {author} {\bibfnamefont {F.}~\bibnamefont {Komissarenko}},
  \bibinfo {author} {\bibfnamefont {D.}~\bibnamefont {Smirnova}}, \bibinfo
  {author} {\bibfnamefont {A.}~\bibnamefont {Vakulenko}}, \bibinfo {author}
  {\bibfnamefont {S.}~\bibnamefont {Kiriushechkina}}, \bibinfo {author}
  {\bibfnamefont {I.}~\bibnamefont {Volkovskaya}}, \bibinfo {author}
  {\bibfnamefont {S.}~\bibnamefont {Guddala}}, \bibinfo {author} {\bibfnamefont
  {V.}~\bibnamefont {Menon}}, \bibinfo {author} {\bibfnamefont
  {A.}~\bibnamefont {Alù}},\ and\ \bibinfo {author} {\bibfnamefont {A.~B.}\
  \bibnamefont {Khanikaev}},\ }\bibfield  {title} {\bibinfo {title} {Photonic
  {Dirac} cavities with spatially varying mass term},\ }\href
  {https://doi.org/10.1126/sciadv.abq4243} {\bibfield  {journal} {\bibinfo
  {journal} {Sci. Adv.}\ }\textbf {\bibinfo {volume} {9}},\ \bibinfo {pages}
  {eabq4243} (\bibinfo {year} {2023})}\BibitemShut {NoStop}%
\bibitem [{rel()}]{relativephase}%
  \BibitemOpen
  \href@noop {} {}\bibinfo {note} {{T}he relative phase of the two spinor
  components is irrelevant for the helicity polarization.}\BibitemShut {Stop}%
\bibitem [{\citenamefont {Greiner}(2000)}]{greiner2000relativistic}%
  \BibitemOpen
  \bibfield  {author} {\bibinfo {author} {\bibfnamefont {W.}~\bibnamefont
  {Greiner}},\ }\href@noop {} {\emph {\bibinfo {title} {Relativistic Quantum
  Mechanics: Wave Equations}}}\ (\bibinfo  {publisher} {Springer, Berlin},\
  \bibinfo {year} {2000})\BibitemShut {NoStop}%
\bibitem [{\citenamefont {Strange}(1998)}]{strange1998relativistic}%
  \BibitemOpen
  \bibfield  {author} {\bibinfo {author} {\bibfnamefont {P.}~\bibnamefont
  {Strange}},\ }\href@noop {} {\emph {\bibinfo {title} {Relativistic Quantum
  Mechanics: With Applications in Condensed Matter and Atomic Physics}}}\
  (\bibinfo  {publisher} {Cambridge University Press, Cambridge},\ \bibinfo
  {year} {1998})\BibitemShut {NoStop}%
\bibitem [{ref()}]{refphi}%
  \BibitemOpen
  \href@noop {} {}\bibinfo {note} {{N}ote that $(\theta,\varphi)$ are spherical
  coordinate parameters in general sense \cite{nguyen2009tunneling}, different
  from the well addressed incident angles $(\alpha,\beta)$ in graphene
  \cite{katsnelson2006chiral,cheianov2006selective} and Weyl semimetal
  \cite{hills2017current}. There is a simple relation between them: for
  $\lambda=1$, $(\theta,\varphi)=(\beta,\alpha)$; and for $\lambda=-1$,
  $(\theta,\varphi)=(\pi-\beta,\alpha-\pi)$, i.e., $\mathbf{p}=\lambda
  p(\sin\beta\cos\alpha,\sin\beta\sin\alpha,\cos\beta)$
  \cite{calogeracos1999history}.}\BibitemShut {Stop}%
\bibitem [{\citenamefont {Bandyopadhyay}\ and\ \citenamefont
  {Cahay}(2008)}]{bandyopadhyay2008introduction}%
  \BibitemOpen
  \bibfield  {author} {\bibinfo {author} {\bibfnamefont {S.}~\bibnamefont
  {Bandyopadhyay}}\ and\ \bibinfo {author} {\bibfnamefont {M.}~\bibnamefont
  {Cahay}},\ }\href@noop {} {\emph {\bibinfo {title} {Introduction to
  Spintronics}}}\ (\bibinfo  {publisher} {CRC press, Boca Raton, FL},\ \bibinfo
  {year} {2008})\BibitemShut {NoStop}%
\bibitem [{\citenamefont {Eschrig}(2015)}]{eschrig2015spin}%
  \BibitemOpen
  \bibfield  {author} {\bibinfo {author} {\bibfnamefont {M.}~\bibnamefont
  {Eschrig}},\ }\bibfield  {title} {\bibinfo {title} {Spin-polarized
  supercurrents for spintronics: a review of current progress},\ }\href@noop {}
  {\bibfield  {journal} {\bibinfo  {journal} {Rep. Prog. Phys.}\ }\textbf
  {\bibinfo {volume} {78}},\ \bibinfo {pages} {104501} (\bibinfo {year}
  {2015})}\BibitemShut {NoStop}%
\bibitem [{\citenamefont {Davies}(1998)}]{davies1998physics}%
  \BibitemOpen
  \bibfield  {author} {\bibinfo {author} {\bibfnamefont {J.~H.}\ \bibnamefont
  {Davies}},\ }\href@noop {} {\emph {\bibinfo {title} {The Physics of
  Low-Dimensional Semiconductors: An Introduction}}}\ (\bibinfo  {publisher}
  {Cambridge University Press, Cambridge},\ \bibinfo {year} {1998})\BibitemShut
  {NoStop}%
\bibitem [{\citenamefont {Markos}\ and\ \citenamefont
  {Soukoulis}(2008)}]{markos2008wave}%
  \BibitemOpen
  \bibfield  {author} {\bibinfo {author} {\bibfnamefont {P.}~\bibnamefont
  {Markos}}\ and\ \bibinfo {author} {\bibfnamefont {C.~M.}\ \bibnamefont
  {Soukoulis}},\ }\href@noop {} {\emph {\bibinfo {title} {Wave Propagation}}}\
  (\bibinfo  {publisher} {Princeton University Press, NJ},\ \bibinfo {year}
  {2008})\BibitemShut {NoStop}%
\bibitem [{\citenamefont {Cot{\u{a}}escu}\ \emph {et~al.}(2007)\citenamefont
  {Cot{\u{a}}escu}, \citenamefont {Gravila},\ and\ \citenamefont
  {Paulescu}}]{cotuaescu2007applying}%
  \BibitemOpen
  \bibfield  {author} {\bibinfo {author} {\bibfnamefont {I.~I.}\ \bibnamefont
  {Cot{\u{a}}escu}}, \bibinfo {author} {\bibfnamefont {P.}~\bibnamefont
  {Gravila}},\ and\ \bibinfo {author} {\bibfnamefont {M.}~\bibnamefont
  {Paulescu}},\ }\bibfield  {title} {\bibinfo {title} {Applying the {D}irac
  equation to derive the transfer matrix for piecewise constant potentials},\
  }\href@noop {} {\bibfield  {journal} {\bibinfo  {journal} {Phys. Lett. A}\
  }\textbf {\bibinfo {volume} {366}},\ \bibinfo {pages} {363} (\bibinfo {year}
  {2007})}\BibitemShut {NoStop}%
\bibitem [{\citenamefont {Sauter}(1931)}]{sauter1931verhalten}%
  \BibitemOpen
  \bibfield  {author} {\bibinfo {author} {\bibfnamefont {F.}~\bibnamefont
  {Sauter}},\ }\bibfield  {title} {\bibinfo {title} {{\"U}ber das {V}erhalten
  eines {E}lektrons im homogenen elektrischen {F}eld nach der relativistischen
  {T}heorie {D}iracs},\ }\href@noop {} {\bibfield  {journal} {\bibinfo
  {journal} {Z. Phys.}\ }\textbf {\bibinfo {volume} {69}},\ \bibinfo {pages}
  {742} (\bibinfo {year} {1931})}\BibitemShut {NoStop}%
\bibitem [{\citenamefont {Klein}\ and\ \citenamefont
  {Nishina}(1929)}]{klein1929streuung}%
  \BibitemOpen
  \bibfield  {author} {\bibinfo {author} {\bibfnamefont {O.}~\bibnamefont
  {Klein}}\ and\ \bibinfo {author} {\bibfnamefont {Y.}~\bibnamefont
  {Nishina}},\ }\bibfield  {title} {\bibinfo {title} {{\"U}ber die {S}treuung
  von {S}trahlung durch freie {E}lektronen nach der neuen relativistischen
  {Q}uantendynamik von {D}irac},\ }\href@noop {} {\bibfield  {journal}
  {\bibinfo  {journal} {Z. Phys.}\ }\textbf {\bibinfo {volume} {52}},\ \bibinfo
  {pages} {853} (\bibinfo {year} {1929})}\BibitemShut {NoStop}%
\bibitem [{\citenamefont {Castro~Neto}\ \emph {et~al.}(2009)\citenamefont
  {Castro~Neto}, \citenamefont {Guinea}, \citenamefont {Peres}, \citenamefont
  {Novoselov},\ and\ \citenamefont {Geim}}]{castro2009electronic}%
  \BibitemOpen
  \bibfield  {author} {\bibinfo {author} {\bibfnamefont {A.~H.}\ \bibnamefont
  {Castro~Neto}}, \bibinfo {author} {\bibfnamefont {F.}~\bibnamefont {Guinea}},
  \bibinfo {author} {\bibfnamefont {N.~M.~R.}\ \bibnamefont {Peres}}, \bibinfo
  {author} {\bibfnamefont {K.~S.}\ \bibnamefont {Novoselov}},\ and\ \bibinfo
  {author} {\bibfnamefont {A.~K.}\ \bibnamefont {Geim}},\ }\bibfield  {title}
  {\bibinfo {title} {The electronic properties of graphene},\ }\href@noop {}
  {\bibfield  {journal} {\bibinfo  {journal} {Rev. Mod. Phys.}\ }\textbf
  {\bibinfo {volume} {81}},\ \bibinfo {pages} {109} (\bibinfo {year}
  {2009})}\BibitemShut {NoStop}%
\bibitem [{\citenamefont {Nguyen}\ and\ \citenamefont
  {Nguyen}(2009)}]{nguyen2009tunneling}%
  \BibitemOpen
  \bibfield  {author} {\bibinfo {author} {\bibfnamefont {H.~C.}\ \bibnamefont
  {Nguyen}}\ and\ \bibinfo {author} {\bibfnamefont {V.~L.}\ \bibnamefont
  {Nguyen}},\ }\bibfield  {title} {\bibinfo {title} {Tunneling of {D}irac
  electrons through one-dimensional potentials in graphene: a {T}-matrix
  approach},\ }\href@noop {} {\bibfield  {journal} {\bibinfo  {journal} {J.
  Phys.: Condens. Matter}\ }\textbf {\bibinfo {volume} {21}},\ \bibinfo {pages}
  {045305} (\bibinfo {year} {2009})}\BibitemShut {NoStop}%
\bibitem [{\citenamefont {Jost}(1957)}]{jost1957bemerkung}%
  \BibitemOpen
  \bibfield  {author} {\bibinfo {author} {\bibfnamefont {R.}~\bibnamefont
  {Jost}},\ }\bibfield  {title} {\bibinfo {title} {Eine {B}emerkung zum {CTP}
  theorem},\ }\href@noop {} {\bibfield  {journal} {\bibinfo  {journal} {Helv.
  Phys. Acta.}\ }\textbf {\bibinfo {volume} {30}},\ \bibinfo {pages} {153}
  (\bibinfo {year} {1957})}\BibitemShut {NoStop}%
\bibitem [{\citenamefont {Bergknoff}\ and\ \citenamefont
  {Thacker}(1979)}]{bergknoff1979structure}%
  \BibitemOpen
  \bibfield  {author} {\bibinfo {author} {\bibfnamefont {H.}~\bibnamefont
  {Bergknoff}}\ and\ \bibinfo {author} {\bibfnamefont {H.~B.}\ \bibnamefont
  {Thacker}},\ }\bibfield  {title} {\bibinfo {title} {Structure and solution of
  the massive {T}hirring model},\ }\href@noop {} {\bibfield  {journal}
  {\bibinfo  {journal} {Phys. Rev. D}\ }\textbf {\bibinfo {volume} {19}},\
  \bibinfo {pages} {3666} (\bibinfo {year} {1979})}\BibitemShut {NoStop}%
\bibitem [{\citenamefont {Thacker}(1981)}]{thacker1981exact}%
  \BibitemOpen
  \bibfield  {author} {\bibinfo {author} {\bibfnamefont {H.~B.}\ \bibnamefont
  {Thacker}},\ }\bibfield  {title} {\bibinfo {title} {Exact integrability in
  quantum field theory and statistical systems},\ }\href@noop {} {\bibfield
  {journal} {\bibinfo  {journal} {Rev. Mod. Phys.}\ }\textbf {\bibinfo {volume}
  {53}},\ \bibinfo {pages} {253} (\bibinfo {year} {1981})}\BibitemShut
  {NoStop}%
\bibitem [{\citenamefont {Froggatt}\ and\ \citenamefont
  {Nielsen}(1991)}]{froggatt1991cpt}%
  \BibitemOpen
  \bibfield  {author} {\bibinfo {author} {\bibfnamefont {C.~D.}\ \bibnamefont
  {Froggatt}}\ and\ \bibinfo {author} {\bibfnamefont {H.~B.}\ \bibnamefont
  {Nielsen}},\ }\href@noop {} {\emph {\bibinfo {title} {Origin of
  Symmetries}}}\ (\bibinfo  {publisher} {World Scientific, Singapore},\
  \bibinfo {year} {1991})\ pp.\ \bibinfo {pages} {86--91}\BibitemShut {NoStop}%
\bibitem [{\citenamefont {Lehnert}(2016)}]{lehnert2016cpt}%
  \BibitemOpen
  \bibfield  {author} {\bibinfo {author} {\bibfnamefont {R.}~\bibnamefont
  {Lehnert}},\ }\bibfield  {title} {\bibinfo {title} {C{PT} symmetry and its
  violation},\ }\href@noop {} {\bibfield  {journal} {\bibinfo  {journal}
  {Symmetry}\ }\textbf {\bibinfo {volume} {8}},\ \bibinfo {pages} {114}
  (\bibinfo {year} {2016})}\BibitemShut {NoStop}%
\bibitem [{\citenamefont {De~Martino}\ \emph {et~al.}(2007)\citenamefont
  {De~Martino}, \citenamefont {Dell’Anna},\ and\ \citenamefont
  {Egger}}]{de2007magnetic}%
  \BibitemOpen
  \bibfield  {author} {\bibinfo {author} {\bibfnamefont {A.}~\bibnamefont
  {De~Martino}}, \bibinfo {author} {\bibfnamefont {L.}~\bibnamefont
  {Dell’Anna}},\ and\ \bibinfo {author} {\bibfnamefont {R.}~\bibnamefont
  {Egger}},\ }\bibfield  {title} {\bibinfo {title} {Magnetic confinement of
  massless {D}irac fermions in graphene},\ }\href@noop {} {\bibfield  {journal}
  {\bibinfo  {journal} {Phys. Rev. Lett.}\ }\textbf {\bibinfo {volume} {98}},\
  \bibinfo {pages} {066802} (\bibinfo {year} {2007})}\BibitemShut {NoStop}%
\bibitem [{\citenamefont {Wu}\ \emph {et~al.}(2010)\citenamefont {Wu},
  \citenamefont {Peeters},\ and\ \citenamefont {Chang}}]{wu2010electron}%
  \BibitemOpen
  \bibfield  {author} {\bibinfo {author} {\bibfnamefont {Z.}~\bibnamefont
  {Wu}}, \bibinfo {author} {\bibfnamefont {F.~M.}\ \bibnamefont {Peeters}},\
  and\ \bibinfo {author} {\bibfnamefont {K.}~\bibnamefont {Chang}},\ }\bibfield
   {title} {\bibinfo {title} {Electron tunneling through double magnetic
  barriers on the surface of a topological insulator},\ }\href@noop {}
  {\bibfield  {journal} {\bibinfo  {journal} {Phys. Rev. B}\ }\textbf {\bibinfo
  {volume} {82}},\ \bibinfo {pages} {115211} (\bibinfo {year}
  {2010})}\BibitemShut {NoStop}%
\bibitem [{\citenamefont {Johnson}\ \emph {et~al.}(1997)\citenamefont
  {Johnson}, \citenamefont {Bennett}, \citenamefont {Yang}, \citenamefont
  {Miller},\ and\ \citenamefont {Shanabrook}}]{johnson1997hybrid}%
  \BibitemOpen
  \bibfield  {author} {\bibinfo {author} {\bibfnamefont {M.}~\bibnamefont
  {Johnson}}, \bibinfo {author} {\bibfnamefont {B.~R.}\ \bibnamefont
  {Bennett}}, \bibinfo {author} {\bibfnamefont {M.~J.}\ \bibnamefont {Yang}},
  \bibinfo {author} {\bibfnamefont {M.~M.}\ \bibnamefont {Miller}},\ and\
  \bibinfo {author} {\bibfnamefont {B.~V.}\ \bibnamefont {Shanabrook}},\
  }\bibfield  {title} {\bibinfo {title} {Hybrid {H}all effect device},\
  }\href@noop {} {\bibfield  {journal} {\bibinfo  {journal} {Appl. Phys.
  Lett.}\ }\textbf {\bibinfo {volume} {71}},\ \bibinfo {pages} {974} (\bibinfo
  {year} {1997})}\BibitemShut {NoStop}%
\bibitem [{\citenamefont {Nogaret}\ \emph {et~al.}(2000)\citenamefont
  {Nogaret}, \citenamefont {Bending},\ and\ \citenamefont
  {Henini}}]{nogaret2000resistance}%
  \BibitemOpen
  \bibfield  {author} {\bibinfo {author} {\bibfnamefont {A.}~\bibnamefont
  {Nogaret}}, \bibinfo {author} {\bibfnamefont {S.~J.}\ \bibnamefont
  {Bending}},\ and\ \bibinfo {author} {\bibfnamefont {M.}~\bibnamefont
  {Henini}},\ }\bibfield  {title} {\bibinfo {title} {Resistance resonance
  effects through magnetic edge states},\ }\href@noop {} {\bibfield  {journal}
  {\bibinfo  {journal} {Phys. Rev. Lett.}\ }\textbf {\bibinfo {volume} {84}},\
  \bibinfo {pages} {2231} (\bibinfo {year} {2000})}\BibitemShut {NoStop}%
\bibitem [{\citenamefont {Goldhaber}(1977)}]{goldhaber1977dirac}%
  \BibitemOpen
  \bibfield  {author} {\bibinfo {author} {\bibfnamefont {A.~S.}\ \bibnamefont
  {Goldhaber}},\ }\bibfield  {title} {\bibinfo {title} {Dirac particle in a
  magnetic field: symmetries and their breaking by monopole singularities},\
  }\href@noop {} {\bibfield  {journal} {\bibinfo  {journal} {Phys. Rev. D}\
  }\textbf {\bibinfo {volume} {16}},\ \bibinfo {pages} {1815} (\bibinfo {year}
  {1977})}\BibitemShut {NoStop}%
\bibitem [{\citenamefont {Andreev}\ and\ \citenamefont
  {Spivak}(2018)}]{andreev2018longitudinal}%
  \BibitemOpen
  \bibfield  {author} {\bibinfo {author} {\bibfnamefont {A.~V.}\ \bibnamefont
  {Andreev}}\ and\ \bibinfo {author} {\bibfnamefont {B.~Z.}\ \bibnamefont
  {Spivak}},\ }\bibfield  {title} {\bibinfo {title} {Longitudinal negative
  magnetoresistance and magnetotransport phenomena in conventional and
  topological conductors},\ }\href@noop {} {\bibfield  {journal} {\bibinfo
  {journal} {Phys. Rev. Lett.}\ }\textbf {\bibinfo {volume} {120}},\ \bibinfo
  {pages} {026601} (\bibinfo {year} {2018})}\BibitemShut {NoStop}%
\bibitem [{\citenamefont {Wang}\ \emph {et~al.}(2021)\citenamefont {Wang},
  \citenamefont {Fu},\ and\ \citenamefont {Shen}}]{wang2021helical}%
  \BibitemOpen
  \bibfield  {author} {\bibinfo {author} {\bibfnamefont {H.-W.}\ \bibnamefont
  {Wang}}, \bibinfo {author} {\bibfnamefont {B.}~\bibnamefont {Fu}},\ and\
  \bibinfo {author} {\bibfnamefont {S.-Q.}\ \bibnamefont {Shen}},\ }\bibfield
  {title} {\bibinfo {title} {Helical symmetry breaking and quantum anomaly in
  massive {D}irac fermions},\ }\href@noop {} {\bibfield  {journal} {\bibinfo
  {journal} {Phys. Rev. B}\ }\textbf {\bibinfo {volume} {104}},\ \bibinfo
  {pages} {L241111} (\bibinfo {year} {2021})}\BibitemShut {NoStop}%
\bibitem [{\citenamefont {Wang}\ \emph {et~al.}(2016)\citenamefont {Wang},
  \citenamefont {Wang}, \citenamefont {Huang},\ and\ \citenamefont
  {Duan}}]{wang2016electronic}%
  \BibitemOpen
  \bibfield  {author} {\bibinfo {author} {\bibfnamefont {J.}~\bibnamefont
  {Wang}}, \bibinfo {author} {\bibfnamefont {N.}~\bibnamefont {Wang}}, \bibinfo
  {author} {\bibfnamefont {H.}~\bibnamefont {Huang}},\ and\ \bibinfo {author}
  {\bibfnamefont {W.}~\bibnamefont {Duan}},\ }\bibfield  {title} {\bibinfo
  {title} {Electronic properties of {S}n{T}e-class topological crystalline
  insulator materials},\ }\href@noop {} {\bibfield  {journal} {\bibinfo
  {journal} {Chin. Phys. B}\ }\textbf {\bibinfo {volume} {25}},\ \bibinfo
  {pages} {117313} (\bibinfo {year} {2016})}\BibitemShut {NoStop}%
\bibitem [{\citenamefont {Gogin}\ \emph
  {et~al.}(2022{\natexlab{a}})\citenamefont {Gogin}, \citenamefont {Rossi},
  \citenamefont {Rossi},\ and\ \citenamefont {Dolcini}}]{Gogin_2022}%
  \BibitemOpen
  \bibfield  {author} {\bibinfo {author} {\bibfnamefont {L.}~\bibnamefont
  {Gogin}}, \bibinfo {author} {\bibfnamefont {L.}~\bibnamefont {Rossi}},
  \bibinfo {author} {\bibfnamefont {F.}~\bibnamefont {Rossi}},\ and\ \bibinfo
  {author} {\bibfnamefont {F.}~\bibnamefont {Dolcini}},\ }\bibfield  {title}
  {\bibinfo {title} {The {D}irac paradox in 1 + 1 dimensions and its
  realization with spin–orbit coupled nanowires},\ }\href
  {https://doi.org/10.1088/1367-2630/ac6cfe} {\bibfield  {journal} {\bibinfo
  {journal} {New J. Phys.}\ }\textbf {\bibinfo {volume} {24}},\ \bibinfo
  {pages} {053045} (\bibinfo {year} {2022}{\natexlab{a}})}\BibitemShut
  {NoStop}%
\bibitem [{\citenamefont {Gogin}\ \emph
  {et~al.}(2022{\natexlab{b}})\citenamefont {Gogin}, \citenamefont {Rossi},\
  and\ \citenamefont {Dolcini}}]{Gogin_2022_2}%
  \BibitemOpen
  \bibfield  {author} {\bibinfo {author} {\bibfnamefont {L.}~\bibnamefont
  {Gogin}}, \bibinfo {author} {\bibfnamefont {F.}~\bibnamefont {Rossi}},\ and\
  \bibinfo {author} {\bibfnamefont {F.}~\bibnamefont {Dolcini}},\ }\bibfield
  {title} {\bibinfo {title} {Electron transport in quantum channels with
  spin–orbit interaction: effects of the sign of the rashba coupling and
  applications to nanowires},\ }\href
  {https://doi.org/10.1088/1367-2630/ac8f66} {\bibfield  {journal} {\bibinfo
  {journal} {New J. Phys.}\ }\textbf {\bibinfo {volume} {24}},\ \bibinfo
  {pages} {093025} (\bibinfo {year} {2022}{\natexlab{b}})}\BibitemShut
  {NoStop}%
\bibitem [{\citenamefont {Cheianov}\ and\ \citenamefont
  {Fal’ko}(2006)}]{cheianov2006selective}%
  \BibitemOpen
  \bibfield  {author} {\bibinfo {author} {\bibfnamefont {V.~V.}\ \bibnamefont
  {Cheianov}}\ and\ \bibinfo {author} {\bibfnamefont {V.~I.}\ \bibnamefont
  {Fal’ko}},\ }\bibfield  {title} {\bibinfo {title} {Selective transmission
  of {D}irac electrons and ballistic magnetoresistance of n-p junctions in
  graphene},\ }\href@noop {} {\bibfield  {journal} {\bibinfo  {journal} {Phys.
  Rev. B}\ }\textbf {\bibinfo {volume} {74}},\ \bibinfo {pages} {041403(R)}
  (\bibinfo {year} {2006})}\BibitemShut {NoStop}%
\bibitem [{\citenamefont {Hills}\ \emph {et~al.}(2017)\citenamefont {Hills},
  \citenamefont {Kusmartseva},\ and\ \citenamefont
  {Kusmartsev}}]{hills2017current}%
  \BibitemOpen
  \bibfield  {author} {\bibinfo {author} {\bibfnamefont {R.~D.~Y.}\
  \bibnamefont {Hills}}, \bibinfo {author} {\bibfnamefont {A.}~\bibnamefont
  {Kusmartseva}},\ and\ \bibinfo {author} {\bibfnamefont {F.~V.}\ \bibnamefont
  {Kusmartsev}},\ }\bibfield  {title} {\bibinfo {title} {Current-voltage
  characteristics of {W}eyl semimetal semiconducting devices, {V}eselago
  lenses, and hyperbolic {D}irac phase},\ }\href@noop {} {\bibfield  {journal}
  {\bibinfo  {journal} {Phys. Rev. B}\ }\textbf {\bibinfo {volume} {95}},\
  \bibinfo {pages} {214103} (\bibinfo {year} {2017})}\BibitemShut {NoStop}%
\bibitem [{\citenamefont {Calogeracos}\ and\ \citenamefont
  {Dombey}(1999)}]{calogeracos1999history}%
  \BibitemOpen
  \bibfield  {author} {\bibinfo {author} {\bibfnamefont {A.}~\bibnamefont
  {Calogeracos}}\ and\ \bibinfo {author} {\bibfnamefont {N.}~\bibnamefont
  {Dombey}},\ }\bibfield  {title} {\bibinfo {title} {History and physics of the
  {K}lein paradox},\ }\href@noop {} {\bibfield  {journal} {\bibinfo  {journal}
  {Contemp. Phys.}\ }\textbf {\bibinfo {volume} {40}},\ \bibinfo {pages} {313}
  (\bibinfo {year} {1999})}\BibitemShut {NoStop}%
\end{thebibliography}%
\end{document}